\documentclass[11pt]{article}
\usepackage{amssymb}
\newcommand{\be}{\begin{equation}}
\newcommand{\ee}{\end{equation}}
\begin{document}
\def\theequation{\arabic{section}.\arabic{equation}}
\begin{titlepage}
\title{Conformally Coupled Inflation}
\author{Valerio Faraoni \\ \\
{\small \it Physics Department and STAR Research Cluster} \\
{\small \it  Bishop's University, 
2600 College St., Sherbrooke, Qu\'{e}bec, Canada J1M~1Z7}}
\date{}
\maketitle   \thispagestyle{empty}  \vspace*{1truecm}

\begin{abstract} 
A massive scalar field in a curved spacetime can propagate along the 
light cone, a causal pathology, which can, in principle, be eliminated 
only if the scalar couples conformally to the Ricci curvature of 
spacetime. This property mandates conformal coupling for the field 
driving inflation in the early universe. During slow-roll inflation, 
this coupling can cause super-acceleration and, as a signature, a blue 
spectrum of primordial gravitational waves.
\end{abstract}
\vspace*{1truecm} 
\begin{center} Keywords: inflation; non-minimal coupling; early universe.
\end{center}     
\end{titlepage}   \clearpage

\section{Introduction}

A period of inflationary expansion of the early universe 
has gradually become to be accepted by most cosmologists as 
a paradigm of the modern scientific picture of the 
universe's history. Although there is no direct proof that 
inflation actually occurred, and it is healthy to 
contemplate alternatives, such as bouncing models 
\cite{bounce1, bounce2}, the 
ekpyrotic universe \cite{ekpyrotic1, ekpyrotic2, 
ekpyrotic3} or string gas cosmology \cite{stringgas1, 
stringgas2, stringgas3}, the temperature anisotropies 
discovered by the {\em COBE
} satellite and further studied 
by the {\em WMAP
} and {\em PLANCK
} missions 
have a spectrum close to the 
Harrison-Zel'dovich 
one predicted by inflation, which 
certainly is some support for the view of an inflationary 
early universe.

Assuming that inflation occurred early on and that it was 
driven by 
some scalar field, $\phi$ (arguably the simplest, although 
not mandatory, class of inflationary scenarios), research 
has for long focused on 
identifying specific scenarios of inflation corresponding 
to particular choices of the scalar field potential, 
$V(\phi)$, motivated by particle physics. Here, we 
argue that the scalar field, $\phi$, driving inflation, should 
be  non-minimally (in fact, conformally) coupled to the 
Ricci curvature of spacetime, $R$, in order to avoid causal 
pathologies. Conformal (or, in general, non-minimal) 
coupling was originally introduced in radiation problems 
\cite{ChernikovTagirov} or in the renormalization of 
scalar fields in curved backgrounds 
\cite{CCJ1,CCJ2,CCJ3,CCJ4,CCJ5,CCJ6,CCJ7}. 
Therefore, it certainly is not obvious that a conventional 
minimally coupled scalar (with timelike or null gradient) 
can suffer from light cone pathologies, but this 
is indeed the case, as was pointed out long ago for test 
fields \cite{SonegoFaraoni}. Let us revisit the argument 
and its consequences for inflation. 

\section{Non-Minimal Coupling}

A scalar field, $\phi$, with mass, $m$, propagating in curved 
spacetime satisfies the Klein-Gordon equation:
\be\label{KleinGordon}
\Box \phi-m^2\phi -\xi R \phi=0 \,
\ee
where the dimensionless non-minimal coupling constant, 
$\xi$, between the scalar and the Ricci curvature is here 
allowed for generality (we will see that minimal coupling, 
corresponding to $\xi=0$, is, in fact, ruled out). 
Here, $\Box=g^{\mu\nu}\nabla_{\mu}\nabla_{\nu}$, 
 where $g_{\mu\nu}$ is the spacetime metric and 
$\nabla_{\mu}$ is its covariant derivative operator. Consider 
the solution of Eq.~(\ref{KleinGordon}) corresponding to a 
delta-like source, which is nothing but the Green function, 
$G_R(x',x)$, of this equation:
\be
\left[ g^{\mu'\nu'} (x') \nabla_{\mu'}\nabla_{\nu'} 
-m^2-\xi R(x') \right] G_R(x',x)=-\delta(x', x) \,
\ee 
where $\delta(x',x)$ is the spacetime delta. By imposing 
the usual boundary conditions, we are restricted to the retarded 
Green function. It is then well known \cite{DeWittBrehme, Friedlander}
 that the retarded Green function, $G_R$, can be 
split as:
\be \label{decomposition}
G_R(x',x)= \Sigma \left( x', x \right) \delta_R\left[ 
\Gamma\left( x', x \right) \right] +W\left( x', x \right)
\Theta\left[ -\Gamma\left( x', x \right)\right] \,
\ee
where $\Gamma\left( x', x \right)$ is the square of the 
proper distance between $x$ and $x'$ calculated along the 
geodesic connecting these two spacetime points (which is 
unique in a normal domain), $\delta_R (\Gamma) $ is the 
usual Dirac delta and $\Theta(-\Gamma)$ is the Heaviside 
step function with support in the past of $x$. The first 
term on the right hand side of Eq.~(\ref{decomposition}) 
describes a contribution to $\phi(x)$ coming from the past 
light cone of $x$, while the second term describes a 
contribution from the interior of this light cone. The 
functions, $\Sigma$ and $W$, are coefficients. 

If the curved spacetime manifold is to be approximated by 
its tangent space (which, loosely speaking, is the spirit 
of the Equivalence Principle of relativity), in the limit, 
$x' \rightarrow x$, in which 
the two points coincide, the Green function must reduce to 
the one of Minkowski space \cite{SonegoFaraoni}, {\em 
i.e.}, it must be: 
\be
\Sigma \left( x', x \right) \rightarrow \Sigma_M \left( x', 
x \right)=\frac{1}{4\pi} \,, \;\;\;\;\;\;\;
W\left( x', x \right) \rightarrow W_M\left( x', x \right) 
\ee
as $x' \rightarrow x$. It is rather straightforward to 
expand all these functions in this limit, obtaining 
\cite{DeWittBrehme, Hadamard, John, SonegoFaraoni}:
\begin{eqnarray}
\Sigma \left( x', x \right) 
&=& \frac{1}{4\pi}+\mbox{O}\left( 
x', x \right) \,\\
&&\nonumber\\
W\left( x', x \right) & = & -\frac{1}{8\pi} \left[ m^2 
+\left( \xi-\frac{1}{6} \right) R(x) \right] 
+\mbox{O}\left( x', x \right) \,\label{6x}\\
&&\nonumber\\
W_M\left( x', x \right) &=& -\frac{m^2}{8\pi} 
+\mbox{O}\left( x', x \right) \,
\end{eqnarray}
where $\mbox{O}\left( x', x \right)$ generically denotes 
terms, which vanish as $x'\rightarrow x$. 
Backscattering of the scalar, $\phi$, can be due to both a 
non-vanishing mass, $m$, or to the background curvature 
appearing in the term, $ - 
\left( \xi-\frac{1}{6} \right) \frac{R(x)}{8\pi}$ in 
Eq.~(\ref{6x}). If 
$m\neq 0$, at spacetime points where 
\begin{equation}
m^2+\left( \xi-\frac{1}{6} \right) R(x) =0 \,
\end{equation}
a {\em massive} scalar, $\phi$, will propagate 
strictly {\em along the light cone}, which is clearly a 
causal pathology. It is even possible to concoct a space of 
constant curvature, $R$, such that the backscattering tail, 
due to the curvature, $\left[ -(\xi-\frac{1}{6}) 
\frac{R(x)}{8\pi} \right]$, exactly compensates the tail, 
$\left[ -\frac{m^2}{8\pi}\right]$, due 
to the mass, $m$. This pathology is possible for $\xi=0$. 
Indeed, the only way to eliminate this disturbing 
possibility is to have $\xi=1/6$ (conformal coupling); then, 
the propagation of a massive $\phi$ is forced to be inside 
the light cone.

Note that conformal invariance has not been imposed or 
implied in any way. It is obtained simply to avoid causal 
pathologies. The physical interpretation of the 
result is the following: because only propagation along 
the light cone is involved in the argument, there must be 
no scale in the physics of the scalar field, which implies 
conformal invariance.

If the argument above applies to a free test 
field, it will also apply to a scalar field in a generic 
potential, $V(\phi)$, and to a gravitating scalar field, which 
always has the previous case as a limit.

Let us review briefly the various formulations of the Equivalence 
Principle. The {\em Weak Equivalence Principle} (WEP) states that if an 
uncharged 
test body is at an initial spacetime point with an initial four-velocity, 
its subsequent trajectory will not depend on its internal structure 
and composition.

The {\em Einstein Equivalence Principle} (EEP) states that (a)~WEP holds; 
(b)~the outcome of any local non-gravitational test experiment is 
independent of the 
velocity of the freely falling apparatus (Local Lorentz Invariance, LLI); 
and (c)~the outcome of any local non-gravitational test experiment is 
independent of where and when in the universe it is performed (Local 
Position Invariance, LPI).

The {\em Strong Equivalence Principle} (SEP) consists of: (a)~WEP holds 
for self-gravitating bodies, as well as for test bodies; (b) the outcome 
of any local test experiment is independent 
of the four-velocity of the freely falling apparatus (Local Lorentz 
Invariance, LLI); and (c)~the outcome of any local test experiment is 
independent of where and when in the universe it is performed (Local 
Position Invariance,~LPI).

The WEP is a statement about mechanics: it requires only the existence of 
preferred trajectories, the free fall trajectories followed by test 
particles, and these curves are the same independently of the 
mass and internal composition of the particles that follow them 
(universality of 
free fall). By itself, WEP does not imply the existence of a 
metric or of geodesic curves (this requirement arises only through the 
EEP by combining the WEP with requirements (b) and (c) \cite{Will}. The EEP 
extends the WEP to all areas of non-gravitational physics. The SEP 
further extends the WEP to self-gravitating bodies and requires 
LLI and LPI to hold also for gravitational experiments, in contrast to the 
EEP. All versions of the Equivalence Principle have been subjected to 
experimental verification, but, thus far, stringent tests only exist for 
the WEP and the EEP \cite{Will}.

Originally \cite{SonegoFaraoni}, the argument for 
$\xi=1/6$ was presented as enforcing the EEP \cite{Will} applied to a test 
or a gravitating field, $\phi$. {\em A posteriori}, however, 
there is no need to invoke the Equivalence Principle, and 
$\phi$ 
could be a gravitational scalar field (for example, in 
a scalar-tensor theory of gravity), about which the 
EEP has nothing to say. Although the argument supporting 
the value, $1/6$, of the coupling constant, $\xi$ (rather 
than the value, $\xi=0$), relies only on the absence of 
causal pathologies in the propagation of $\phi$-waves, it 
is interesting to elaborate on it in light of the 
recent paper \cite{Gerard} on theories of gravity 
satisfying the SEP. The author of \cite{Gerard} 
looks for ways to implement the SEP on theories of gravity and, on the
 basis of the analogy with the Standard Model of particle physics, 
concludes that the SEP is embodied by the condition on 
the Riemann tensor: 
\be\label{SEP}
\nabla_{\sigma} {R^{\sigma}}_{\lambda\mu\nu}=0 \,
\ee
which is analogous to the condition: 
\be\label{YM}
D_{\mu} F^{\mu\nu}=0
\ee
for non-Abelian Yang-Mills fields of strength, $F^{\mu\nu}$, 
which satisfy $\left[ D_{\mu}, D_{\nu} \right]=iF_{\mu\nu}$ 
(where $D_{\mu}$ is the covariant derivative). The Riemann 
tensor satisfies the analogous relation:
\be
{\left[ \nabla_{\mu}, \nabla_{\nu} 
\right]^{\alpha}}_{\beta}= -{R^{\alpha}}_{\beta\mu\nu} \,
\ee
(This characterization of the 
SEP, however, is different from the traditional one of, {e.g.}, 
\cite{Will}, presented above.) 
Eq.~(\ref{SEP}) expresses the condition that ``gravitons 
gravitate the same way that gluons glue'' \cite{Gerard}. 
Consider general 
scalar-tensor theories of gravity described by the 
(Jordan frame) action: 
\be
S_{ST}=\frac{1}{16\pi} \int d^4x \sqrt{-g} \left[ \phi 
R-\frac{\omega(\phi)}{\phi} \, g^{\mu\nu}\nabla_{\mu}\phi
\nabla_{\nu}\phi \right] +S^{(matter)}\,
\ee
where the Brans-Dicke-like $\phi$ is of gravitational 
nature (we use units in which 
Newton's constant, $G$, and the speed of light, $c$, are 
unity and the 
Brans-Dicke coupling, $\omega(\phi)$, is a 
function of $\phi$). In general, the gravitational or 
non-gravitational nature of a field depends on the 
conformal frame representation of the theory; see the 
discussion in \cite{ThomasStefanoValerio}. In short, 
scalar-tensor gravity can be discussed in the 
Jordan frame (meaning the set of variables, $\left( g_{\mu\nu}, \phi \right)$), 
in which the scalar field, $\phi$, couples explicitly to
the Ricci curvature and matter is minimally coupled (which has the consequence
that massive test particles follow timelike geodesics). Alternatively,
one can describe the theory in the Einstein conformal frame, the set of 
variables, $\left( \tilde{g}_{\mu\nu}, \tilde{\phi} \right)$, related to the 
Jordan frame by the conformal redefinition of the metric:
\be
g_{\mu\nu} \longrightarrow \tilde{g}_{\mu\nu}= \phi \, g_{\mu\nu}
\ee
 and the non-linear field redefinition:
\be
d\tilde{\phi} = \sqrt{\frac{2\omega(\phi )+3}{16\pi} } \, 
\frac{d\phi}{\phi} \,
\ee
In the Einstein frame, the scalar field has canonical kinetic energy and couples
minimally to gravity ({\em i.e.}, there is no explicit coupling between $\phi$
and $R$), but it couples directly to the the matter Lagrangian in the action. 
As a consequence, uncharged particles in the Einstein frame do not follow 
geodesics of the metric, $\tilde{g}_{\mu\nu}$, but deviate from them, due to 
a force proportional to the gradient of the scalar field. Massless particles,
the physics of which is conformally invariant, follow null geodesics in 
both frames ({e.g.}, \cite{mybook}).

It turns out that imposing the SEP condition (\ref{SEP}) selects 
only two possible theories \cite{Gerard}. These are Nordstrom's scalar 
gravity (in which the metric is conformally flat and there 
is only a scalar degree of freedom) and the theory with: 
\be
\omega(\phi) = \frac{3\phi}{2\left( \phi -1 \right)} \,
\ee
In the latter case, the field redefinition, $\phi 
\rightarrow \varphi$, with: 
\be
\phi = 1-\frac{4\pi \varphi^2}{3}
\ee
recasts the action as:
\be
S=\frac{1}{16 \pi } \int d^4x \sqrt{-g}\left[ \left( 
\frac{1}{2}- 
\frac{\varphi^2}{12}\right)R -\frac{1}{2} \, 
g^{\mu\nu}\nabla_{\mu}\varphi\nabla_{\nu}\varphi 
\right]+S^{(matter)} \,
\ee
which is the action for a conformally coupled scalar field. 
In other words, insisting that the gravitational Brans-Dicke-like 
scalar field $\phi$ satisfies the EEP 
(or that the theory satisfies the SEP), leads to the requirement 
that it be conformally coupled. The traditional SEP 
amounts to imposing that the Weak Equivalence Principle 
of mechanics is 
satisfied also by gravitating bodies, plus local Lorentz 
invariance and local position invariance \cite{Will}. 
Following the definition of SEP adopted in \cite{Gerard}, it would 
seem that the SEP would correspond to imposing the EEP also 
on gravitational fields.

Now, if $\phi$ is a gravitational scalar field in a theory 
of gravity alternative to general relativity, there is no 
reason for it to satisfy the EEP. Moreover, the 
Brans-Dicke-like field of scalar-tensor gravity is not 
supposed to be the one driving inflation---even in the 
extended and hyperextended inflationary scenarios based on 
Brans-Dicke gravity and on more general scalar-tensor 
theories, respectively; 
it is a second  non-gravitational scalar field that is 
responsible for inflation (see, {e.g.}, the review in 
\cite{mybook}). However, any field satisfying 
Eq.~(\ref{KleinGordon}) should be conformally 
coupled, $\xi=1/6$. Let us review the consequences of 
conformal coupling if $\phi$ is the scalar field driving 
inflation in the early universe.

\section{Consequences for Inflation}

It is well known that, if one quantizes a scalar field on a 
curved background, a non-minimal coupling to the Ricci scalar, 
$R$, is 
introduced, even if it was absent in the classical theory 
\cite{CCJ1,CCJ2,CCJ3,CCJ4,CCJ5,CCJ6,CCJ7}. In asymptotically free grand 
unified theories, depending 
on the gauge group and the matter content, $\xi$ is a 
running coupling and, generically, $1/6$ is a stable 
infrared fixed point \cite{Odintsov1,Odintsov2,Odintsov3,Odintsov4,Odintsov5,Odintsov6,Odintsov7,Odintsov8}. 
According to the 
previous (classical) argument, the inflation field fueling 
inflation should be 
coupled conformally. Then, one should revisit inflation, 
keeping in mind that conformal coupling is not an option, 
but is required for consistency of the theory. 
Over the years, several authors have studied non-minimally 
coupled inflatons, usually in a rather opportunistic way, 
{\em 
i.e.}, the coupling constant, $\xi$, was usually considered 
as a free parameter to be adjusted at will in order to 
alleviate fine-tuning problems in the potential 
\cite{NMCinflation1,NMCinflation2,NMCinflation3,NMCinflation4,NMCinflation5,NMCinflation6,NMCinflation7,NMCinflation8,NMCinflation9,NMCinflation10,NMCinflation11,NMCinflation12,NMCinflation13,NMCinflation14,NMCinflation15,NMCinflation16,NMCinflation17}. Now the 
value of $\xi$ is forced upon us. It has been 
demonstrated that viable scenarios of inflation for an 
unperturbed universe can occur with non-minimal coupling 
\cite{NMCinflation1,NMCinflation2,NMCinflation3,NMCinflation4,NMCinflation5,NMCinflation6,NMCinflation7,NMCinflation8,NMCinflation9,NMCinflation10,NMCinflation11,NMCinflation12,NMCinflation13,NMCinflation14,NMCinflation15,NMCinflation16,NMCinflation17}. A possible 
obstacle is the fact that the 
effective term, 
$-\xi R\phi^2/2$, in the Lagrangian could, in principle, spoil 
the flatness of an inflationary potential, $V(\phi)$ 
\cite{NMCinflation1,NMCinflation2,NMCinflation3,NMCinflation4,NMCinflation5,NMCinflation6,NMCinflation7,NMCinflation8,NMCinflation9,NMCinflation10,NMCinflation11,NMCinflation12,NMCinflation13,NMCinflation14,NMCinflation15,NMCinflation16,NMCinflation17}, but this 
difficulty is not crucial. What is more, new features of 
the dynamics emerge, which are not possible when $\xi=0$ 
\cite{CQGBrussels, AnnPhys}. By 
adopting a spatially flat 
Friedmann-Lem\^aitre-Robertson-Walker metric:
\be
ds^2=-dt^2+a^2(t)\left( dx^2+dy^2+dz^2 \right) \,
\ee
the field equations are: 
\begin{eqnarray}
&&H^2=\frac{\kappa}{3} \, \rho \, \label{efe1}\\
&&\nonumber\\
&& \frac{\ddot{a}}{a} = \dot{H}+H^2 = -\frac{\kappa}{6} \left( \rho+P 
\right) \,\label{efe2}\\
&&\nonumber\\
&&\ddot{\phi}+3H\dot{\phi}+\frac{dV}{d\phi}+\xi R\phi=0 \,
\end{eqnarray}
where $\rho $ and $P$ are the energy density and pressure 
of the cosmic fluid, respectively, $\kappa \equiv 8\pi 
G$ ($G$ being Newton's constant) and an overdot denotes 
differentiation with respect to the comoving time, $t$. 
Eqs.~(\ref{efe1}) and (\ref{efe2}) yield:
\be\label{cazzocubo}
\dot{H}=-\frac{\kappa}{2} \left( \rho +P \right) 
\ee
and, therefore, $P<-\rho$ (a ``phantom'' equation of state) 
is equivalent to $\dot{H}>0$. A regime with $\dot{H}>0$, due 
to non-minimal coupling, called {\em superinflation}, was 
studied already in the 1980s \cite{LucchinMatarrese85a, 
LucchinMatarrese85b}. 
Minimally coupled scalar 
fields have $\rho=\frac{\dot{\phi}^2}{2}+V(\phi) $ and 
$ P=\frac{\dot{\phi}^2}{2}-V(\phi) $; hence, the derivative 
$\dot{H}$ in Eq.~(\ref{cazzocubo}) gives: 
\be
\dot{H}=-\kappa\dot{\phi}^2/2 \leq 0 \,
\ee 
By contrast, for a non-minimally coupled scalar field, it 
is:
\begin{eqnarray}
\rho &=& \frac{\dot{\phi}^2}{2}+V(\phi) +3\xi H\phi \left( 
H\phi +2\dot{\phi} \right) \,\\
&&\nonumber\\
P &=& \frac{\dot{\phi}^2}{2} - V(\phi) -\xi \left[ 4H\phi 
\dot{\phi} + 2 \dot{\phi}^2 +2 \phi \ddot{\phi} +\left( 
2 \dot{H} + 3H^2 \right) \phi^2 \right]\,
\end{eqnarray} 
and $\dot{H}>0$ is a possibility. Indeed, exact solutions 
exhibiting explicitly this super-acceleration have been 
found in the context of early universe inflation 
\cite{integrability, IJMPD} and of present-day quintessence 
\cite{BigSmash}.
 
Technically speaking, the non-minimally coupled scalar 
field action:
\be\label{action}
S_{NMC}=\int d^4x \sqrt{-g} \left[ \left( 
\frac{1}{2\kappa}-\frac{\xi}{2} \phi^2 \right)R-\frac{1}{2} 
\, \nabla^{\mu}\phi \nabla_{\mu}\phi -V(\phi) \right]
\ee
is a scalar-tensor action, and gauge-independent 
formalisms have been developed to study cosmological 
perturbations in this class of theories. 
One can fix a gauge and proceed to study perturbations 
in that gauge, but there is the risk that the results are unphysical, 
an artifact of pure gauge modes. Indeed, the gauge-dependence 
problem plagued the early studies of cosmological perturbations 
produced during inflation. An alternative is to identify 
gauge-invariant variables and derive equations for 
these gauge-invariant quantities that assume the same 
form in all gauges. When this is done, a gauge-invariant 
formalism is obtained, which has the advantage of being 
completely gauge-independent and the disadvantage that the 
gauge-invariant variables are not physically transparent---they can receive a physical interpretation once a gauge is fixed. 
The original gauge-invariant formalism, due to Bardeen \cite{Bardeen}, 
has been refined over the years and was designed for cosmology in 
the context of general relativity. Here, we adopt the 
Bardeen-Ellis-Bruni-Hwang formalism 
\cite{Bardeen,EllisBruni1,EllisBruni2,EllisBruni3}, 
that is, a version of the Bardeen formalism \cite{Bardeen}, 
refined by Ellis, Bruni 
and Hwang and adapted by Hwang to a wide class of theories 
of gravity alternative to general relativity \cite{Hwang1,Hwang2,Hwang3,Hwang4,Hwang5,Hwang6,Hwang8}. 
In fact, non-minimally coupled scalar field theory is a special 
case of scalar-tensor gravity, as can be seen by tracing in 
reverse the path outlined in Section~2, and it is straightforward to
apply Hwang's formalism to this theory. The application of this formalism 
to non-minimally coupled 
inflation was reviewed in \cite{think}. Slow-roll inflation 
with de Sitter universes as attractors in phase space is 
possible. 

There are four slow-roll parameters, as opposed to the two 
of minimally coupled inflation (for comparison, $-\epsilon_1$ and 
$-\epsilon_2$ coincide with the usual parameters, $\epsilon$ 
and $\eta$, of minimally coupled  inflation) \cite{Hwang1,Hwang2,Hwang3,Hwang4,Hwang5,Hwang6,Hwang8,think}: 
\begin{eqnarray} \epsilon_1 &=& \frac{\dot{H}}{H^2} \,, 
\;\;\;\;\;\;\;
\epsilon_2 = \frac{\ddot{\phi}}{H \dot{\phi}} \,\\
&&\nonumber\\
\epsilon_3 &=& - \frac{\xi \kappa \phi \dot{\phi}}{H 
\left[ 1-\left( \frac{\phi}{\phi_1} \right)^2 \right]} 
\,, \;\;\;\;\;\;\; \epsilon_4 = - \frac{\xi (1-6\xi) 
\kappa \phi \dot{\phi}}{H 
\left[ 1-\left( \frac{\phi}{\phi_2} \right)^2 \right]} 
\end{eqnarray}
(with $\phi_{1,2}$ constants), and $\epsilon_4$ vanishes for 
$\xi=1/6$. The 
spectral indices of scalar and tensor 
perturbations in the slow-roll approximation are then 
\cite{Hwang1,Hwang2,Hwang3,Hwang4,Hwang5,Hwang6,Hwang8}:
\begin{eqnarray}
n_S &=& 1+2\left( 2\epsilon_1 -\epsilon_2 +\epsilon_3 
\right)\,\\
&&\nonumber\\
n_T &=& 2\left( 2\epsilon_1-\epsilon_3 \right) \,
\end{eqnarray}
There is a possible signature of conformal coupling in the 
cosmic microwave background sky. In an inflationary 
super-acceleration regime, $\dot{H}>0$ (which is impossible 
with minimal coupling and realistic scalar field 
potentials), it is $\epsilon_1>0$, and one 
can obtain a {\em blue} spectrum of gravitational waves, 
$n_T>0$. 
Blue spectra of tensor perturbations are impossible in the 
standard scenarios of inflation (they are, however, possible in 
certain non-inflationary scenarios) with $\xi=0$ (for 
which 
$n_T=4\dot{H}/H 
\leq 0$). More power is shifted to small wavelengths in 
comparison with minimally coupled inflation, 
which is interesting for the gravitational wave community, 
because it increases the chance of detecting cosmological 
gravitational waves with future space-based 
interferometers. 

\section{Conclusions}

Supporting the idea that the inflaton field is conformally, 
rather than minimally, coupled is actually a pretty 
conservative view. Not doing so means allowing for a 
possible pathology in the local propagation of the 
inflaton, {\em i.e.}, the possibility that this field 
propagates along the light cone when it is 
massive. This problem is even more serious during inflation 
because, in slow-roll, the cosmic dynamics are close to a 
de 
Sitter attractor for which $R$ is constant, and one could 
even have the causal pathology mentioned above (or be 
very close to it) at {\em 
every} spacetime point. This would indeed be a radical 
departure from known physics, which cannot be justified. The 
only way out of this conundrum is if $\xi=1/6$, and 
inflationary scenarios should be adapted to this 
constraint.

\section*{Acknowledgments} It is a pleasure to thank Sebastiano Sonego 
for his leading role in the investigation of non-minimal coupling long 
ago, which led to the development of this project. We thank also three 
referees for useful remarks. This work is supported by the Natural 
Sciences and Engineering Research Council of Canada.

\end{document}